\documentclass[twoside,hidelinks,11pt]{article}

\usepackage{jmlr2e}
\usepackage{hyperref}
\usepackage{listings}
\usepackage{natbib}
\usepackage{courier}

\ShortHeadings{June 15, 2015, Postprint for JMLR.}{Heaton}
\firstpageno{1}

\begin{document}

\title{Encog: Library of Interchangeable Machine Learning Models for Java and C\#}

\author{\name Jeff Heaton \email jeffheaton@acm.org \\
       \addr Graduate School of Computer and Information Sciences\\
       Nova Southeastern University\\
       Fort Lauderdale, FL 33314, USA}

\editor{Cheng Soon Ong}

\maketitle

\begin{abstract}
This paper introduces the Encog library for Java and C\#, a scalable, adaptable, multi-platform machine learning framework that was first released in 2008.  Encog allows a variety of machine learning models to be applied to datasets using regression, classification, and clustering.  Various supported machine learning models can be used interchangeably with minimal recoding.  Encog uses efficient multithreaded code to reduce training time by exploiting modern multicore processors.  The current version of Encog can be downloaded from \url{http://www.encog.org}.
\end{abstract}


\section{Intention and Goals}

This paper describes the Encog API for Java and C\# that is provided as a JAR or DLL library.  The C\# version of Encog is also compatible with the Xamarin Mono package.  Encog has an active community that has provided many enhancements that are beyond the scope of this paper.  This includes extensions such as Javascript, GPU processing, C/C++ support, Scala support, and interfaces to various automated trading platforms.  The scope of this paper is limited to the Java and C\# API.

Encog allows the Java or C\# programmer to experiment with a wide range of machine language models using a simple, consistent interface for clustering, regression, and classifications.  This allows the programmer to construct applications that discover which model provides the most suitable fit for the data.  Encog provides basic tools for automated model selection.  Most Encog models are implemented as efficient multithreaded algorithms to reduce processing time.  This often allows Encog to perform more efficiently than many other Java and C\# libraries, as demonstrated empirically by \citet{taheretaheri:2010:Online} and \citet{matviykiv2012data}.  \citet{DBLP:conf/sisy/IuhaszMN13} and \citet{DBLP:journals/jms/Ramos-PollanGO12} also saw favorable results when evaluating Encog to similar libraries.

The Encog's API is presented in an intuitive object-oriented paradigm that allows various models, optimization algorithms, and training algorithms to be highly interchangeable.  However, beneath the API, the models are represented as one and two-dimensional arrays.  This internal representation allows for highly efficient calculation.  The API shields the programmer from the complexity of model calculation and fitting. 

Encog contains nearly 400 unit tests to ensure consistency between the Java and C\# model implementations.  Expected results are calculated and cross-checked between the two platforms.  A custom pseudorandom number generator (PRNG) is used in both language's unit tests to ensure that even stochastic models produce consistent, verifiable test results. 

Encog contains nearly 150 examples to demonstrate the use of the API in a variety of scenarios. These examples include simple prediction, time series, simulation, financial applications, path finding, curve fitting, and other applications.  Documentation for Encog is provided as Java/C\# docs and an online wiki.  Additionally, discussion groups and a Stack Overflow tag are maintained for support.  Links to all of these resources can be found at \url{http://www.encog.org}.  

\section{Framework Overview}
The design goal of Encog is to provide interchangeable models with efficient, internal implementations.  The Encog framework supports machine learning models with multiple training algorithms.  These models are listed here:

\begin{itemize}
\itemsep0em 
\item Adaline, Feedforward, Hopfield, PNN/GRNN, RBF \& NEAT neural networks
\item generalized linear regression (GLM)
\item genetic programming (tree-based)
\item k-means clustering
\item k-nearest neighbors
\item linear regression
\item self-organizing map (SOM)
\item simple recurrent network (Elman and Jordan)
\item support vector machine (SVM)
\end{itemize}

Encog provides optimization algorithms such as particle swarm optimization (PSO) \citep{Poli08}, genetic algorithms (GA), Nelder-Mead and simulated annealing.  These algorithms can optimize a vector to minimize a loss function; consequently, these algorithms can fit model parameters to datasets.  

Propagation-training algorithms for neural network fitting, such as back propagation \citep{Rumelhart1988}, resilient propagation \citep{Riedmiller92rprop}, Levenberg-Marquardt \citep{marquardt:1963}, quickpropagation \citep{Fahlman88}, and scaled conjugate gradient \citep{Moller93ascaled} are included.  Neural network pruning and model selection can be used to find optimal network architectures.  Neural network architectures can be automatically built by a genetic algorithm using NEAT and HyperNEAT \citep{Stanley2002}.      

A number of preprocessing tools are built into the Encog library.  Collected data can be divided into training, test, and validation sets.  Time-series data can be encoded into data windows.  Quantitative data can be normalized by range or z-score to prevent biases in some models.  \citet{Masters1993} normalizes qualitative data using  one-of-n encoding or equilateral encoding. Encog uses these normalization techniques.

Encog also contains extensive support for genetic programming using a tree representation \citep{koza1993}.  A full set of mathematical and programming functions are provided.  Additionally, new functions can be defined.  Constant nodes can either be drawn from a constant pool or generated as needed.  Rules can optionally be added to simplify expressions and penalize specific genome patterns. 

\section{API Overview}
One of the central design philosophies of Encog is to allow models to be quickly interchanged without a great deal of code modification. A classification example will demonstrate this interchangeability, using the iris dataset \citep{fisher36lda}. Portions of this classification example are presented in this paper, using the Java programming language.  The complete example, in both Java and C\#, is provided in the \textit{Encog Quick Start Guide} (available from \url{http://www.encog.org}.  The Quick Start Guide also provides regression and time-series examples.

The following example learns to predict the species of an iris flower by using four types of measurements from each flower. To begin, the program loads the iris dataset's CSV file.  In addition to CSV, Encog contains classes to read fixed-length text, JDBC, ODBC, and XML data sources.  The iris dataset is loaded, and the four measurement columns are defined as continuous values.
\begin{lstlisting}
VersatileDataSource source = new CSVDataSource(irisFile, false,
    CSVFormat.DECIMAL_POINT);
VersatileMLDataSet data = new VersatileMLDataSet(source);
data.defineSourceColumn("sepal-length", 0, ColumnType.continuous);
data.defineSourceColumn("sepal-width", 1, ColumnType.continuous);
data.defineSourceColumn("petal-length", 2, ColumnType.continuous);
data.defineSourceColumn("petal-width", 3, ColumnType.continuous);
\end{lstlisting}

The species of Iris is defined as nominal value.  Defining the columns as continuous, nominal or ordinal allows Encog to determine the appropriate way to encode these data for a model.  For specialized cases, it is possible to override Encog's encoding defaults for any model type. 
\begin{lstlisting}			
ColumnDefinition outputColumn = data.defineSourceColumn("species", 4,
    ColumnType.nominal);
\end{lstlisting}
Once the columns have been defined, the file is analyzed to determine minimum, maximum, and other statistical properties of the columns.  This allows the columns to be properly normalized and encoded by Encog for modeling.
\begin{lstlisting}			
data.analyze();
data.defineSingleOutputOthersInput(outputColumn);
\end{lstlisting}
Next the model type is defined to be a feedforward neural network. 
\begin{lstlisting}				
EncogModel model = new EncogModel(data);
model.selectMethod(data, MLMethodFactory.TYPE_FEEDFORWARD);
\end{lstlisting}
Only the above line needs to be changed to switch to model types that include the following:
\begin{itemize}
\itemsep0em 
\item \textbf{MLMethodFactory.SVM}:  support vector machine
\item \textbf{MLMethodFactory.TYPE\_RBFNETWORK}: RBF neural network
\item \textbf{MLMethodFactor.TYPE\_NEAT}: NEAT neural network
\item \textbf{MLMethodFactor.TYPE\_PNN}: probabilistic neural network
\end{itemize}

Next the dataset is normalized and encoded.  Encog will automatically determine the correct normalization type based on the model chosen in the last step.  For model validation, 30\% of the data are held back.  Though the validation sampling is random, a seed of 1001 is used so that the items selected for validation remain constant between program runs. Finally, the default training type is selected.
\begin{lstlisting}		
data.normalize();
model.holdBackValidation(0.3, true, 1001);
model.selectTrainingType(data);
\end{lstlisting}			

The example trains using a 5-fold cross-validated technique that chooses the model with the best validation score. The resulting training and validation errors are displayed.
\begin{lstlisting}	
MLRegression bestMethod = (MLRegression)model.crossvalidate(5, true);
System.out.println( "Training error: " + EncogUtility.calculateRegressionError(bestMethod, model.getTrainingDataset()));
System.out.println( "Validation error: " + EncogUtility.calculateRegressionError(bestMethod, model.getValidationDataset()));
\end{lstlisting}
Display normalization parameters and final model.	
\begin{lstlisting}	
NormalizationHelper helper = data.getNormHelper();
System.out.println(helper.toString());
System.out.println("Final model: " + bestMethod);
\end{lstlisting}

\section{Future Plans and Conclusions}

A number of enhancements are planned for Encog.  Gradient boosting machines (GBM) and deep learning are two future model additions.  Several planned enhancements will provide interoperability with other machine learning packages.  Future versions of Encog will have the ability to read and write Weka Attribute-Relation File Format (ARFF) and libsvm data files. Encog will gain the ability to load and save models in the Predictive Model Markup Language (PMML) format.  A code contribution by \citet{Mosca:mscdissertation} will soon be integrated, enhancing Encog's ensemble learning capabilities.


\acks{The Encog community has been very helpful for bug reports, bug fixes, and feature suggestions.  Contributors to Encog include Olivier Guiglionda, Seema Singh, C{\'{e}}sar Roberto de Souza, and others.  A complete list of contributors to Encog can be found at the GitHub repository: \url{https://github.com/encog}.  Alan Mosca, Department of Computer Science 
and Information Systems, Birkbeck, University of London, UK, created Encog's ensemble functionality.  Matthew Dean, Marc Fletcher and Edmund Owen, Semiconductor Physics Research Group, University of Cambridge, UK, created Encog's RBF Neural network model.
}


\newpage

\vskip 0.2in
\bibliography{heaton14b}

\begin{thebibliography}{15}
\providecommand{\natexlab}[1]{#1}
\providecommand{\url}[1]{\texttt{#1}}
\expandafter\ifx\csname urlstyle\endcsname\relax
  \providecommand{\doi}[1]{doi: #1}\else
  \providecommand{\doi}{doi: \begingroup \urlstyle{rm}\Url}\fi

\bibitem[Fahlman(1988)]{Fahlman88}
S.~Fahlman.
\newblock An empirical study of learning speed in back-propagation networks.
\newblock Technical report, Carnegie Mellon University, 1988.

\bibitem[Fisher(1936)]{fisher36lda}
R.~Fisher.
\newblock The use of multiple measurements in taxonomic problems.
\newblock \emph{Annals of Eugenics}, 7\penalty0 (7):\penalty0 179--188, 1936.

\bibitem[Koza(1993)]{koza1993}
J.~Koza.
\newblock \emph{Genetic programming - on the programming of computers by means
  of natural selection}.
\newblock Complex adaptive systems. MIT Press, 1993.
\newblock ISBN 978-0-262-11170-6.

\bibitem[Luhasz et~al.(2013)Luhasz, Munteanu, and
  Negru]{DBLP:conf/sisy/IuhaszMN13}
G.~Luhasz, V.~Munteanu, and V.~Negru.
\newblock Data mining considerations for knowledge acquisition in real time
  strategy games.
\newblock In \emph{{IEEE} 11th International Symposium on Intelligent Systems
  and Informatics, {SISY} 2013, Subotica, Serbia, September 26-28, 2013}, pages
  331--336, 2013.
\newblock \doi{10.1109/SISY.2013.6662596}.

\bibitem[Marquardt(1963)]{marquardt:1963}
D.~Marquardt.
\newblock An algorithm for least-squares estimation of nonlinear parameters.
\newblock \emph{SIAM Journal on Applied Mathematics}, 11\penalty0 (2):\penalty0
  431--441, 1963.
\newblock \doi{10.1137/0111030}.

\bibitem[Masters(1993)]{Masters1993}
T.~Masters.
\newblock \emph{Practical Neural Network Recipes in C++}.
\newblock Academic Press Professional, Inc., San Diego, CA, USA, 1993.
\newblock ISBN 0-12-479040-2.

\bibitem[Matviykiv and Faitas(2012)]{matviykiv2012data}
O.~Matviykiv and O.~Faitas.
\newblock Data classification of spectrum analysis using neural network.
\newblock \emph{Lviv Polytechnic National University}, 2012.

\bibitem[M{\o}ller(1993)]{Moller93ascaled}
M.~M{\o}ller.
\newblock A scaled conjugate gradient algorithm for fast supervised learning.
\newblock \emph{NEURAL NETWORKS}, 6\penalty0 (4):\penalty0 525--533, 1993.

\bibitem[Mosca(2012)]{Mosca:mscdissertation}
A.~Mosca.
\newblock Extending encog: A study on classifier ensemble techniques.
\newblock Master's thesis, Birkbeck, University of London, 2012.

\bibitem[Poli(2008)]{Poli08}
R.~Poli.
\newblock Analysis of the publications on the applications of particle swarm
  optimisation.
\newblock \emph{J. Artif. Evol. App.}, 2008:\penalty0 4:1--4:10, January 2008.
\newblock ISSN 1687-6229.

\bibitem[Ramos{-}Poll{\'{a}}n et~al.(2012)Ramos{-}Poll{\'{a}}n,
  Guevara{-}L{\'{o}}pez, and Oliveira]{DBLP:journals/jms/Ramos-PollanGO12}
Ra{\'{u}}l Ramos{-}Poll{\'{a}}n, Miguel~{\'{A}}ngel Guevara{-}L{\'{o}}pez, and
  Eug{\'{e}}nio~C. Oliveira.
\newblock A software framework for building biomedical machine learning
  classifiers through grid computing resources.
\newblock \emph{J. Medical Systems}, 36\penalty0 (4):\penalty0 2245--2257,
  2012.
\newblock \doi{10.1007/s10916-011-9692-3}.

\bibitem[Riedmiller and Braun(1992)]{Riedmiller92rprop}
M.~Riedmiller and H.~Braun.
\newblock Rprop - a fast adaptive learning algorithm.
\newblock Technical report, Proc. of ISCIS VII), Universitat, 1992.

\bibitem[Rumelhart et~al.(1988)Rumelhart, Hinton, and Williams]{Rumelhart1988}
D.~Rumelhart, G.~Hinton, and R.~Williams.
\newblock Neurocomputing: Foundations of research.
\newblock In James~A. Anderson and Edward Rosenfeld, editors,
  \emph{Neurocomputing: Foundations of Research}, chapter Learning
  Representations by Back-propagating Errors, pages 696--699. MIT Press,
  Cambridge, MA, USA, 1988.
\newblock ISBN 0-262-01097-6.

\bibitem[Stanley and Miikkulainen(2002)]{Stanley2002}
K.~Stanley and R.~Miikkulainen.
\newblock Evolving neural networks through augmenting topologies.
\newblock \emph{Evol. Comput.}, 10\penalty0 (2):\penalty0 99--127, June 2002.
\newblock ISSN 1063-6560.

\bibitem[Taheri(2014)]{taheretaheri:2010:Online}
T.~Taheri.
\newblock Benchmarking and comparing encog, neuroph and joone neural networks.
\newblock \url{http://goo.gl/A56iyx}, June 2014.
\newblock Accessed: 2014-10-9.

\end{thebibliography}

\end{document}